\documentclass[graybox]{svmult}
\usepackage{natbib}

\usepackage{mathptmx}       % selects Times Roman as basic font
\usepackage{helvet}         % selects Helvetica as sans-serif font
\usepackage{courier}        % selects Courier as typewriter font
\usepackage{type1cm}        % activate if the above 3 fonts are
\usepackage{makeidx}         % allows index generation
\usepackage{graphicx}        % standard LaTeX graphics tool
\usepackage{multicol}        % used for the two-column index
\usepackage[bottom]{footmisc}% places footnotes at page bottom
\makeindex             % used for the subject index
\begin{document}

\title*{Four-winged flapping flyer in forward flight}
\titlerunning{Four-winged flapping flyer}
\author{R. Godoy-Diana\inst{1,*}, P. Jain\inst{1,2}, M. Centeno\inst{1,3}, A. Weinreb\inst{1} \and B. Thiria\inst{1}}
\authorrunning{Godoy-Diana \emph{et al.}} 
\institute{$^1${Physique et M\'ecanique des Milieux Het\'erog\`enes (PMMH), CNRS UMR 7636; ESPCI ParisTech; UPMC (Paris 6); Univ. Paris Diderot (Paris 7), 10 rue Vauquelin, 75231 Paris, Cedex 5, France}\\
$^2$Indian Institute of Technology, Kharagpur, India\\
$^3$Facultad de Ciencias, Universidad Nacional Aut\'onoma de M\'exico\\
$^*$Corresponding author: ramiro@pmmh.espci.fr}

\thispagestyle{empty} \maketitle \thispagestyle{empty}
\setcounter{page}{1}

%\pacs{
%    87.85.gj    % Biomechanics. Movement and locomotion
%}

\abstract{
We study experimentally a four-winged flapping flyer with chord-wise flexible wings in a self-propelled setup. For a given physical configuration of the flyer (i.e. fixed distance between the forewing and hindwing pairs and fixed wing flexibility), we explore the kinematic parameter space constituted by the flapping frequency and the forewing-hindwing phase lag. Cruising speed and consumed electric power measurements are performed for each point in the $(f,\varphi)$ parameter space and allow us to discuss the problem of performance and efficiency in four-winged flapping flight. We show that different phase-lags are needed for the system to be optimised for fastest flight or lowest energy consumption. A conjecture of the underlying mechanism is proposed in terms of the coupled dynamics of the forewing-hindwing phase lag and the deformation kinematics of the flexible wings.
}

%\maketitle

\section{Introduction}
\noindent Flapping flyers display an extremely rich variety of maneuvers because of the multiple kinematic parameters that rule the unsteady production of aerodynamic forces. From a biological point of view, the case of four-winged flyers capable of out-of-phase motion between forewings and hindwings such as dragonflies is particularly interesting. In the words of \cite{azuma1985}: ``Dragonflies can hover, fly at high speed and maneuver skillfully in the air in order to defend their territory, feed on live prey and mate in tandem formation''. Their body and wing kinematics have been studied extensively \cite[][]{alexander1984,alexander1986,azuma1988,ruppell1989,wakeling1997part2,wakeling1997part3} and flow visualizations in tethered- and free-flight configurations have demonstrated the crucial role of unsteady mechanisms such as the formation of leading-edge vortices in the production of lift \cite[][]{thomas2004}. Forewing-hindwing phase-lag has been shown in hovering configurations to be determinant for flight performance \cite[][]{maybury2004}: optimal efficiencies have been found for out-of-phase beating whereas in-phase motion of forewings and hindwings has been shown to produce stronger force \cite[][]{wang2007,usherwood2008}. The physical mechanisms behind these differences in performance have nonetheless not yet been completely elucidated, and open questions remain in particular when considering the role of wing elasticity. Wing deformation is important because it can passively modify the effective angle of attack of a flapping wing, determining thus its  force production dynamics. 

In the present paper we address this problem experimentally using a four-winged self-propelled model mounted on a ``merry-go-round''. The setup is a modified version of the one used by \cite{thiria2010} and \cite{ramananarivo2011}, where the thrust force produced by the wings makes the flyer turn around a central axis. A constant cruising speed is achieved for a given wingbeat frequency when the thrust generated is balanced by the net aerodynamic drag on the flyer. These previous works have shown that passive mechanisms associated to wing flexibility govern the flying performance of a flapping wing flyer with chord-wise flexible wings. These determine for instance that the elastic nature of the wings can lead not only to a substantial reduction of the consumed power, but also to an increment of the propulsive force. Here we introduce a new parameter using a model with two pairs of wings. In addition to the flapping frequency and wing flexibility, the thrust production is now also determined by the phase lag $\varphi$ between the forewings and the hindwings. 
  
\begin{figure}
\centering
\includegraphics[width=\linewidth]{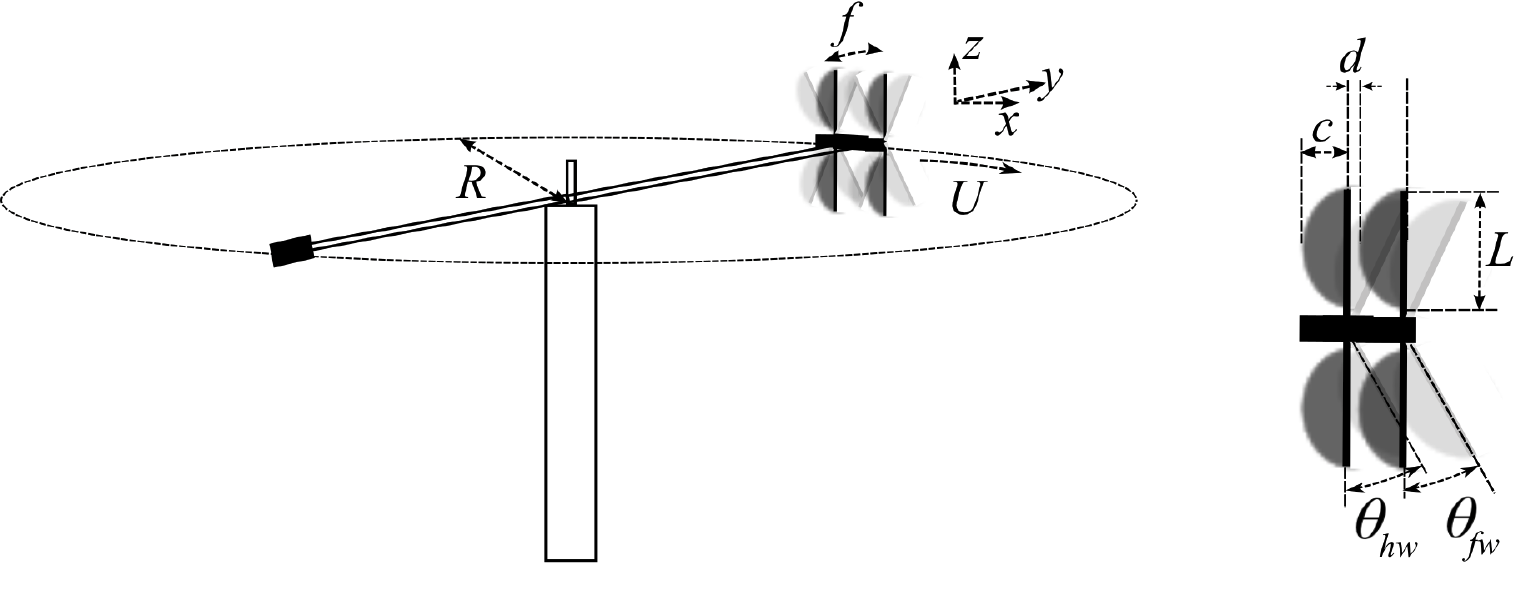}
\includegraphics[height=0.35\linewidth]{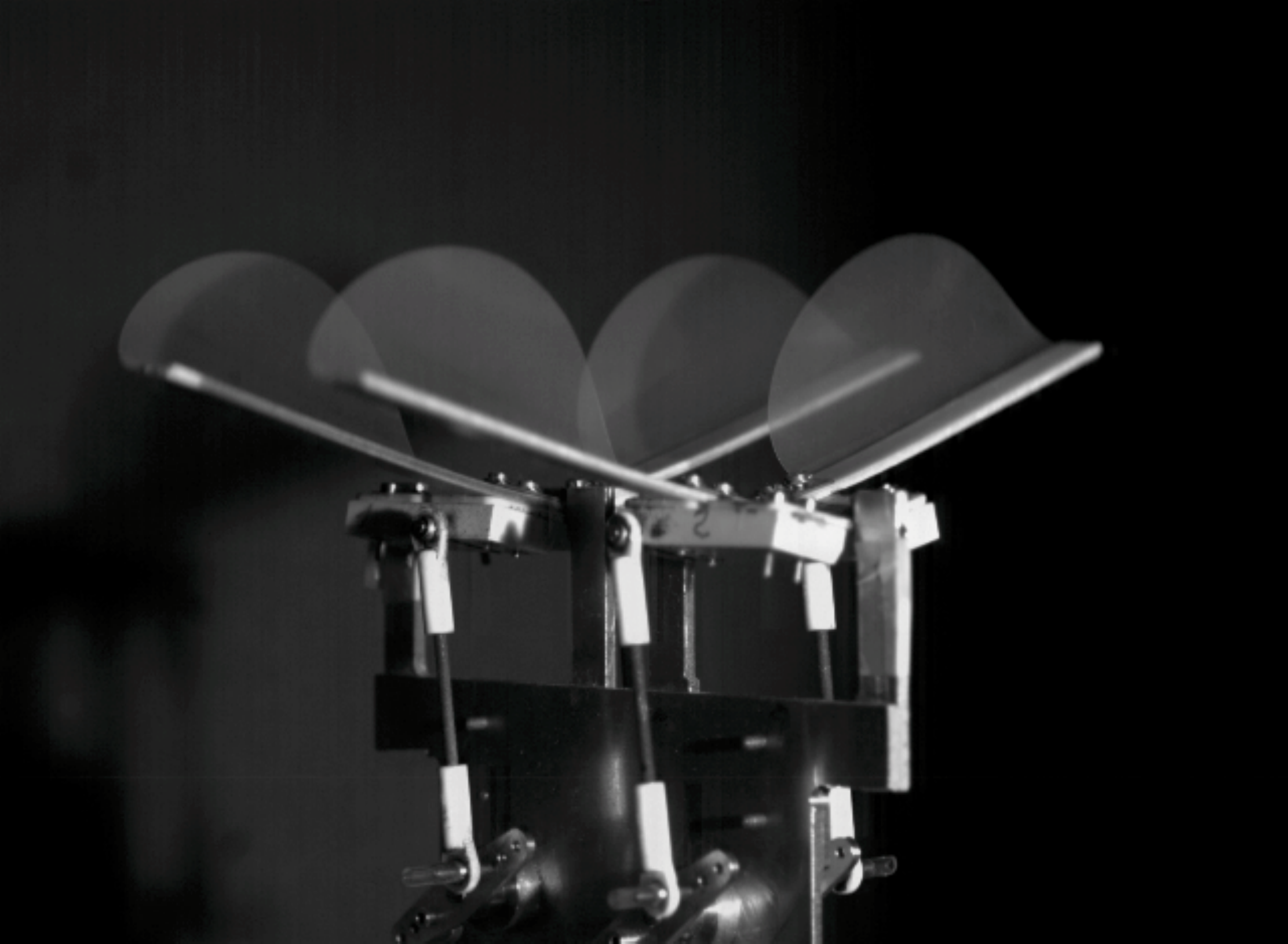}
\includegraphics[height=0.35\linewidth]{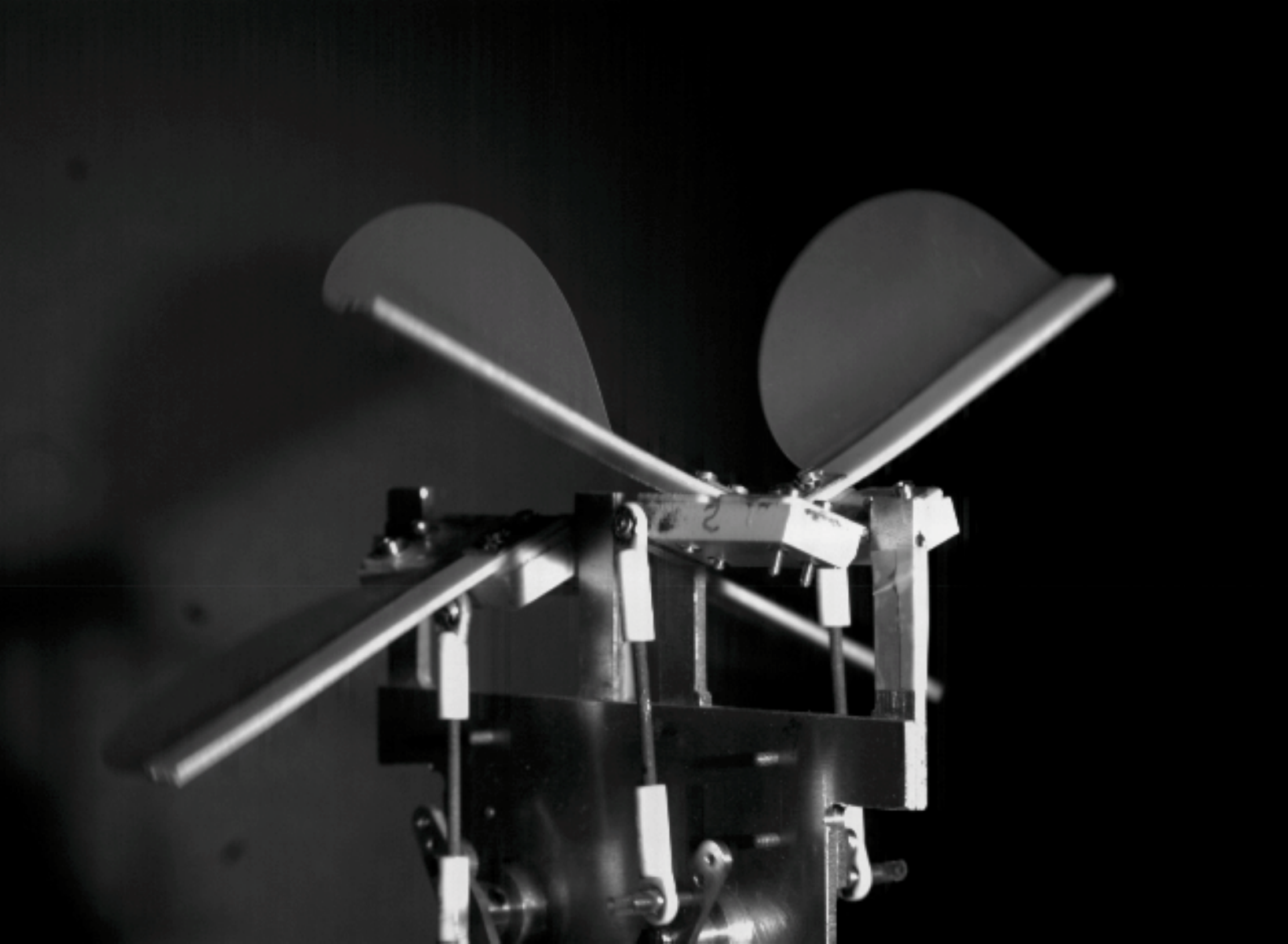}
  \caption{Top: sketch of the experimental setup. Bottom: photos of the flapping flyer with forewing-hindwing phase lags 0 (left) and $\pi$ (right). Each wing has a semi-circular shape so that $c=30$mm and $L=60$mm. The distance between the wings $d=1$ mm, so that the trailing edge of the forewing and the leading edge of the hindwing almost touch when the wings are aligned without bending; the stroke amplitude $\max(\theta_{fw})=\max(\theta_{hw})=\theta_0=37^{\circ}$. Wings are made of 0.05mm thick Mylar that gives a flexural rigidity $B=3.3\times 10^{-5}$ N.m. The leading edge is thicker (1mm) and made of fiberglass so that it can be considered rigid and the deformation exclusively chord-wise.}
\label{figura}
\end{figure}

\section{Experimental setup}

Figure \ref{figura} shows a sketch of the experimental setup. As in \cite{thiria2010}, the stroke plane is parallel to the shaft linking the flyer to the central bearing of the wheel. In addition to the four-winged instead of two-winged flyer, the setup also differs from \cite{thiria2010} in that a counterweight has been added using an opposite radial shaft to balance the system. The fluctuating lift force is thus directed radially and absorbed by the shaft. The two wings are driven by a single direct-current motor with a set of gears that allows to fix the phase difference between the forewings and the hindwings. All wings beat thus at the same frequency which was varied between 15 and 30 Hz. We have reduced the parameter space in the experiments reported here by fixing the physical characteristics of the flyer. Namely, the distance between the wings $d$, the stroke amplitude $\theta_0$ and the chord-wise flexural rigidity of the wings. It should of course be noted that these parameters in the present tandem wing configuration, in particular the wing spacing $d$, should in general be analysed simultaneously with the forewing-hindwing phase lag \cite[][]{maybury2004,rival2011}. The motion of the wings is described using the angles of the forewing and hindwing leading edges to the $xz$-plane, $\theta_{fw}$ and $\theta_{hw}$, respectively (see Fig.~\ref{figura}), as 

\begin{equation}
\theta_{fw}=\theta_0\sin(2\pi f t) \;\;  \mathrm{and} \;\; \theta_{hw}=\theta_0\sin(2\pi f t -\varphi)\;,
\end{equation}
 where $f$ is the flapping frequency and the phase lag $\varphi$ is varied between 0 and $2\pi$. For $0<\varphi<\pi$ the forewing is leading whereas for $\pi<\varphi<2\pi$ it is the hindwing that leads. The Reynolds number $\mathrm{Re}=Uc/\nu$ based on the cruising speed and the chord length was in the range of 1000 to 4000.

The measured quantities are the cruising flight speed $U$ and the consumed power $P_i$. In the study of the two winged flyer of \cite{thiria2010} and \cite{ramananarivo2011} an additional independent setup was used to measure the thrust force $F$ by holding the flyer in a stationary position. The product of the force and cruising speed measurements was then an estimate of the aerodynamic power $P_a=FU$. One of the disadvantages of that procedure was that the two measurements did not correspond to the same flight configuration: while $U$ corresponds to self-propelled cruising flight, $F$ measured at a fixed station corresponds to a "hovering" flight  configuration. Here we avoid that problem by using an estimate of aerodynamic power obtained only from the velocity measurements, as explained in the next section.

\section{Results}
\begin{figure}[t]
\centering
\includegraphics[width=0.49\linewidth]{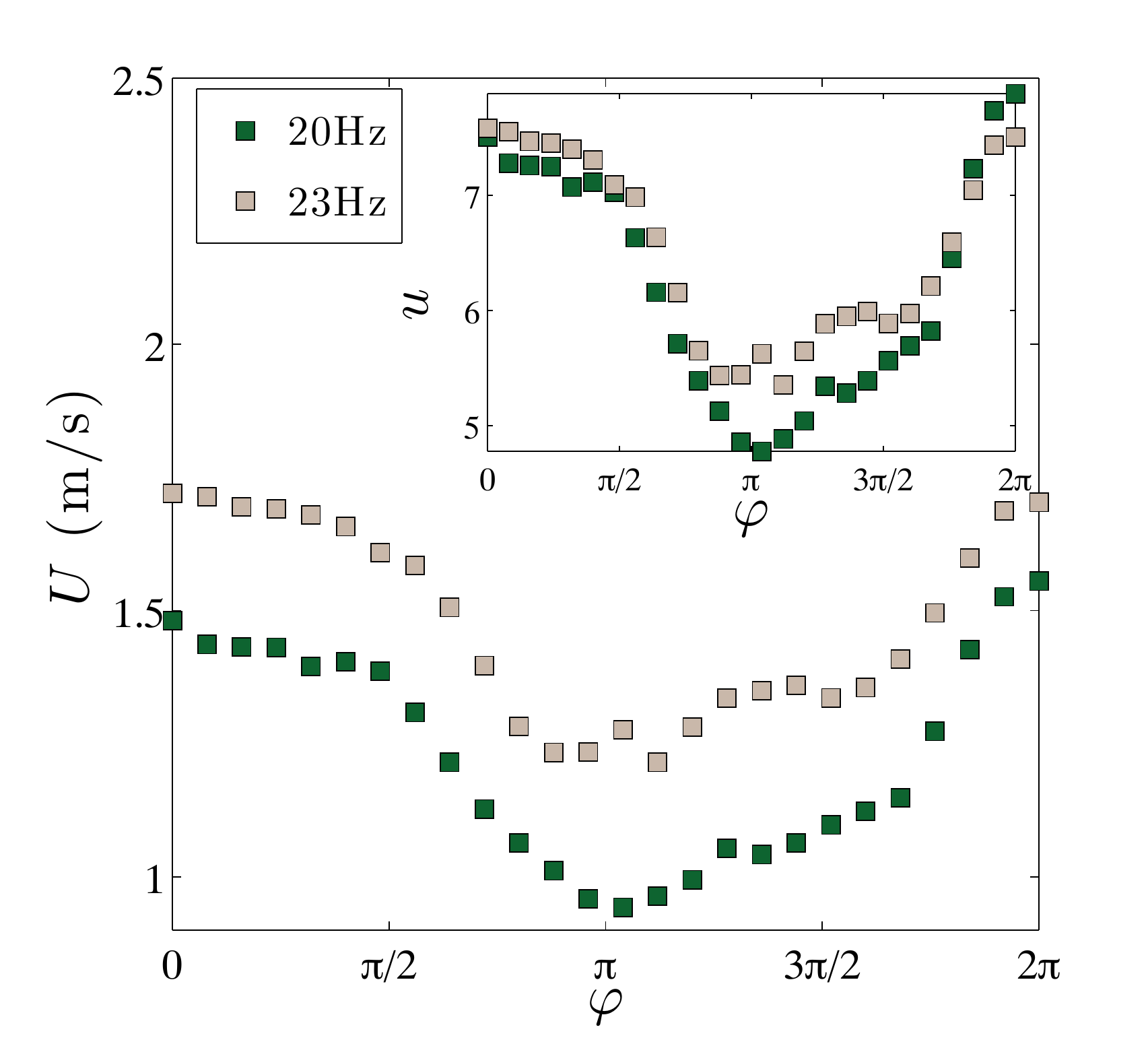}
\includegraphics[width=0.49\linewidth]{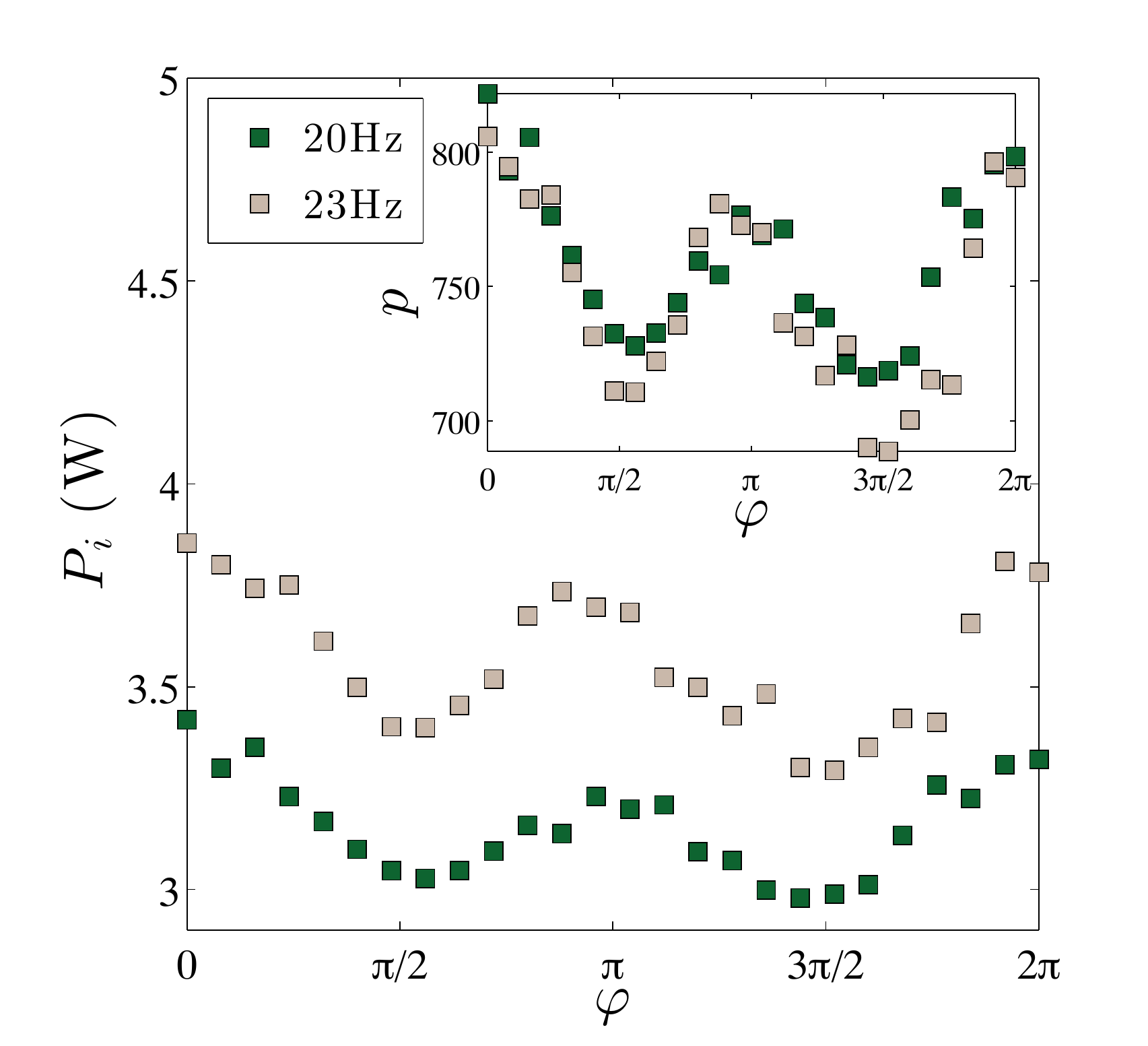}
\caption{Cruising speed and consumed power as a function of the forewing-hindwing phase lag for two different flapping frequencies.}
\label{fig_results}
\end{figure}

\textbf{(i) Flying performance.}
The aerodynamic interactions are thus ruled by $\varphi$ and their effect can be directly measured in the performance parameters of the experiment: the cruising speed $U$ and the consumed power $P_i$. In order to get a clear picture of the effect of each parameter we show first in Fig.~\ref{fig_results} two data series corresponding to different  flapping frequencies, plotting $U$ and $P_i$ as a function of $\varphi$. It can be seen that the phasing between the wings produces a net effect in performance, the fastest cruising flight velocities corresponding to a range around in-phase flapping ($\varphi=0$), but the picture becomes richer when looking at the consumed power. Indeed, the latter has two peaks around $\varphi\approx0$ and $\pi$.  The previous series lie in a regime where  increasing the flapping frequency shifts the curves to higher flying speeds and higher consumed power. The observed trend is readily explained using the non-dimensional expressions $p=P_i c/B\omega_f$ and $u=U/A_0\omega_f$ defined by \cite{ramananarivo2011}, where $A_0$ is the amplitude of oscillation of the leading edge at mid span given by $A_0=(L/2)\sin\theta_0$ and $\omega_f=2\pi f$. The insets in Fig.~\ref{fig_results} show the behaviour of the dimensionless quantities.

The increase of speed with increasing flapping frequency is not indefinite however, an effect shown in Fig.~\ref{fig_contours}, where $U$ is plotted in coloured contours in a $(\varphi,f)$-space for  $\varphi\in[0,\pi]$ and scanning the full range of flapping frequencies available experimentally. For all phase lags, a clear maximum of the attained cruising speed occurs always around 24Hz. We will discuss in the following that this optimal frequency is related to the elastic properties of the wings. The second plot in Fig.~\ref{fig_contours} shows the consumed power $P_i$ in the same parameter space. Here the main observation is that, while not surprisingly consumed power increases monotonically with increasing flapping frequency, the effect of the phase lag on energy expenditure shown previously in Fig.~\ref{fig_results} is clearly present regardless the flapping frequency. Because of this effect, for different phase lags the ranges of frequency explored changed, giving for instance a maximum frequency for $\varphi=0$ of around 30Hz while at $\varphi\in[\pi/2,3\pi/4]$ the frequency could reach 35Hz. We use those two measurements to define the following expression of efficiency, considering that the aerodynamic thrust power is proportional to $U^3$ (velocity times thrust force, the latter being $\sim U^2$):
\begin{figure}[t]
\centering
\includegraphics[width=\linewidth]{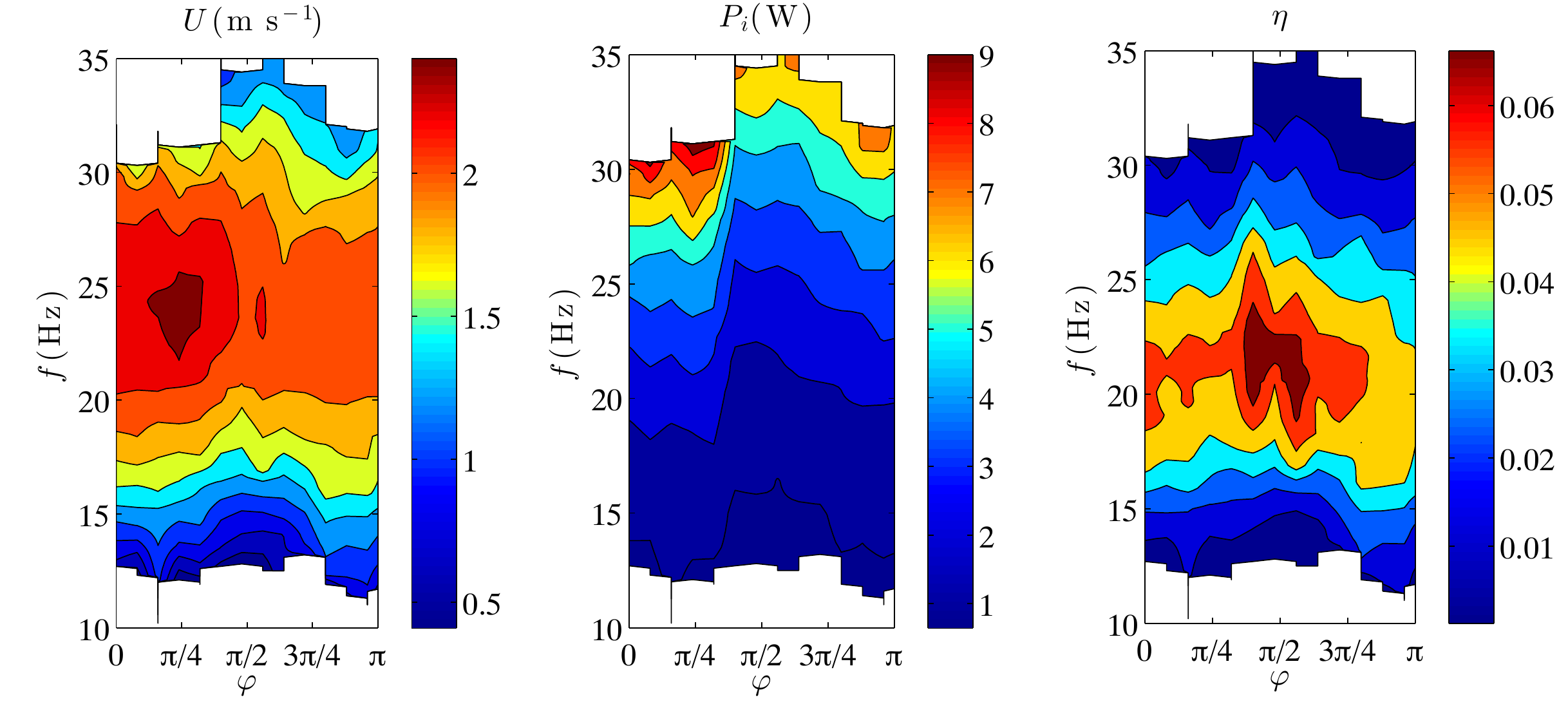}
\caption{Cruising speed, consumed power and efficiency (see text) as a function of the forewing-hindwing phase lag and the flapping frequency. Only the first half o the phase-lag $\varphi$ range where the forewing is leading was examined here.}
\label{fig_contours}
\end{figure}

\begin{equation}
\eta=\frac{\frac 1 2 \rho S U^3}{P_i}
\label{eff}
\end{equation}

\noindent where $S$ is the effective wing surface. Other definitions of efficiency using purely dynamical parameters \cite[e.g.][]{kang2011} should give equivalent results to the expression \ref{eff} chosen here in terms of the measured consumed power $P_i$.
 It can be seen that the optimum in terms of efficiency is shifted toward larger phase lags (around $\varphi\approx\pi/2$) than the optimum in terms of maximum cruising speed.  

\begin{figure}[p]
\centering
\includegraphics[width=\linewidth]{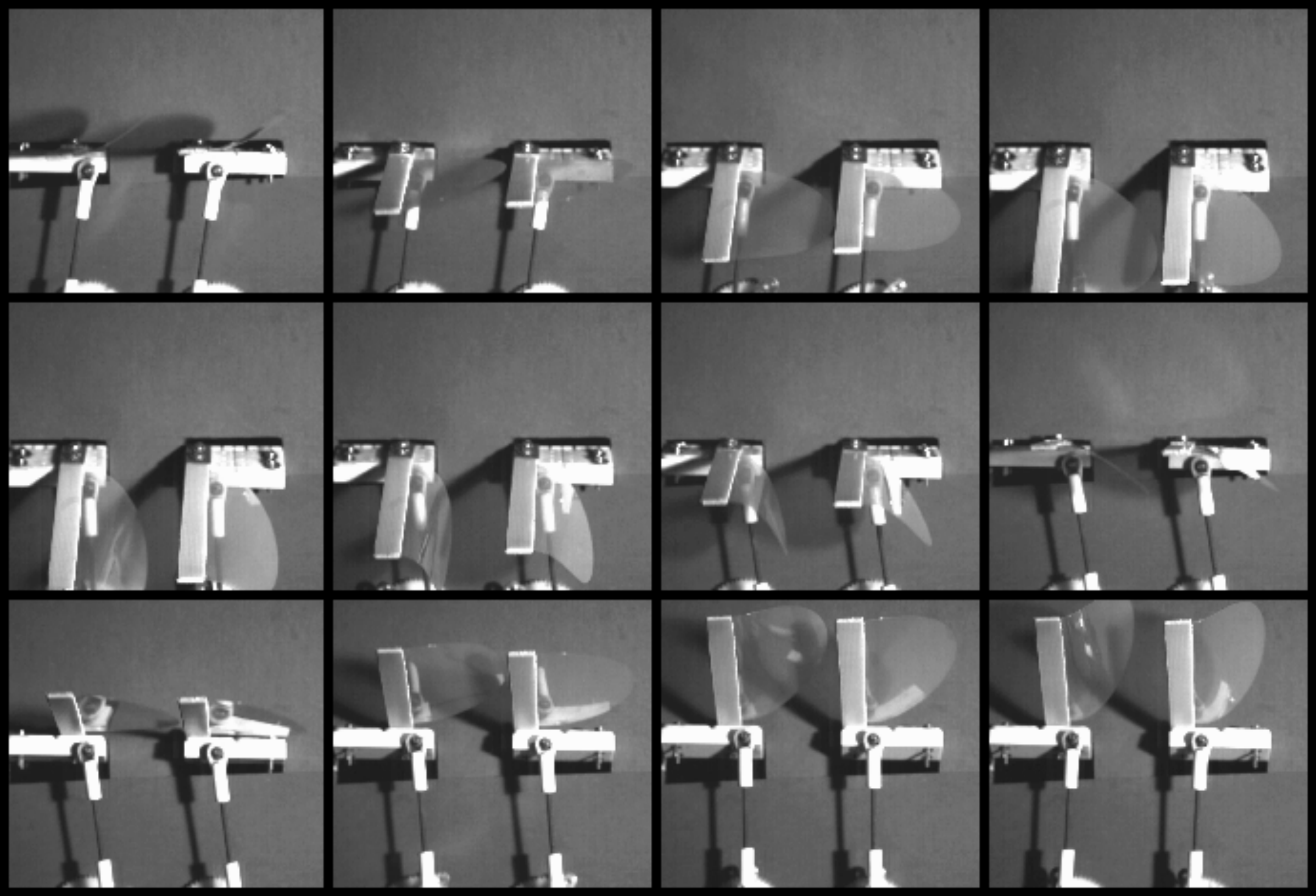}
  \caption{Sequence showing the kinematics of the flapping wings at $f=23$Hz and $\varphi=0$}
\label{fig_phi0}
\end{figure}

\begin{figure}[p]
\centering
\includegraphics[width=\linewidth]{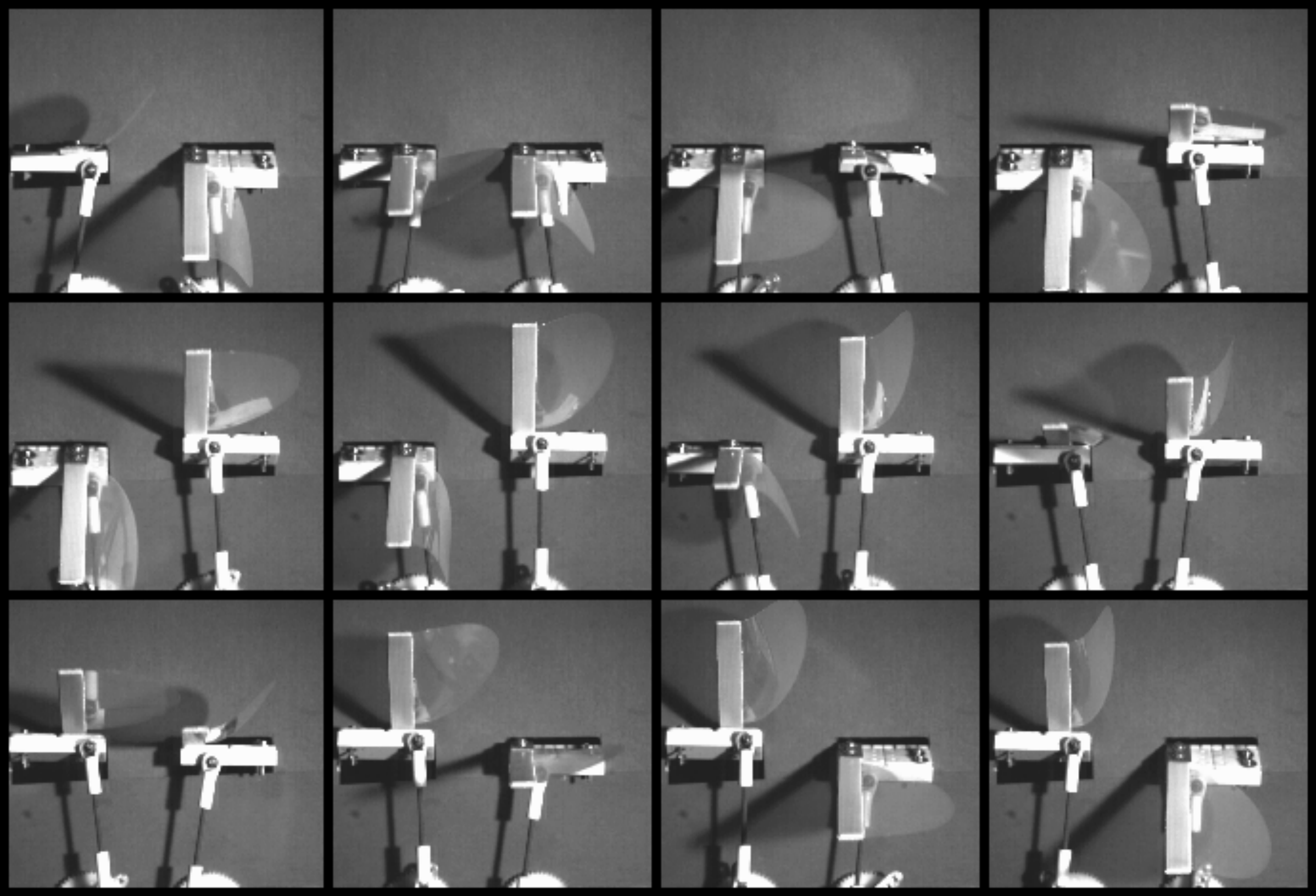}
  \caption{Sequence showing the kinematics of the flapping wings at $f=23$Hz and $\varphi=1.3\pi$}
\label{fig_phi1}
\end{figure}

\bigskip
\noindent\textbf{(ii) Wing kinematics.} In addition to the performance measurements, the wing motion was recorded using high-speed video in a fixed frame. That is, with the mechanical insect not mounted on the merry-go-round but on a fixed base. Figures \ref{fig_phi0} and \ref{fig_phi1} show typical time series for $\varphi=0$ and $1.3\pi$, respectively. A dark screen was used to mask the right side wings and have a clearer view of the kinematics. The main feature that can be seen in these time series of snapshots is the large deformation of the wings during the flapping cycle. Indeed, the wings bend over a length scale that is of the same order of magnitude than the chord length. In addition to the kinematics of the compliant wings, which can be followed for each wing independently, figures  \ref{fig_phi0} and \ref{fig_phi1} hint on the complex interaction that arises from the combination of the bending dynamics and the imposed forewing-hindwing phase lag. We will analyse this interaction in the following by tracking the motion of different points on each wing. 

\begin{figure}[t]
\centering
\includegraphics[width=1\linewidth]{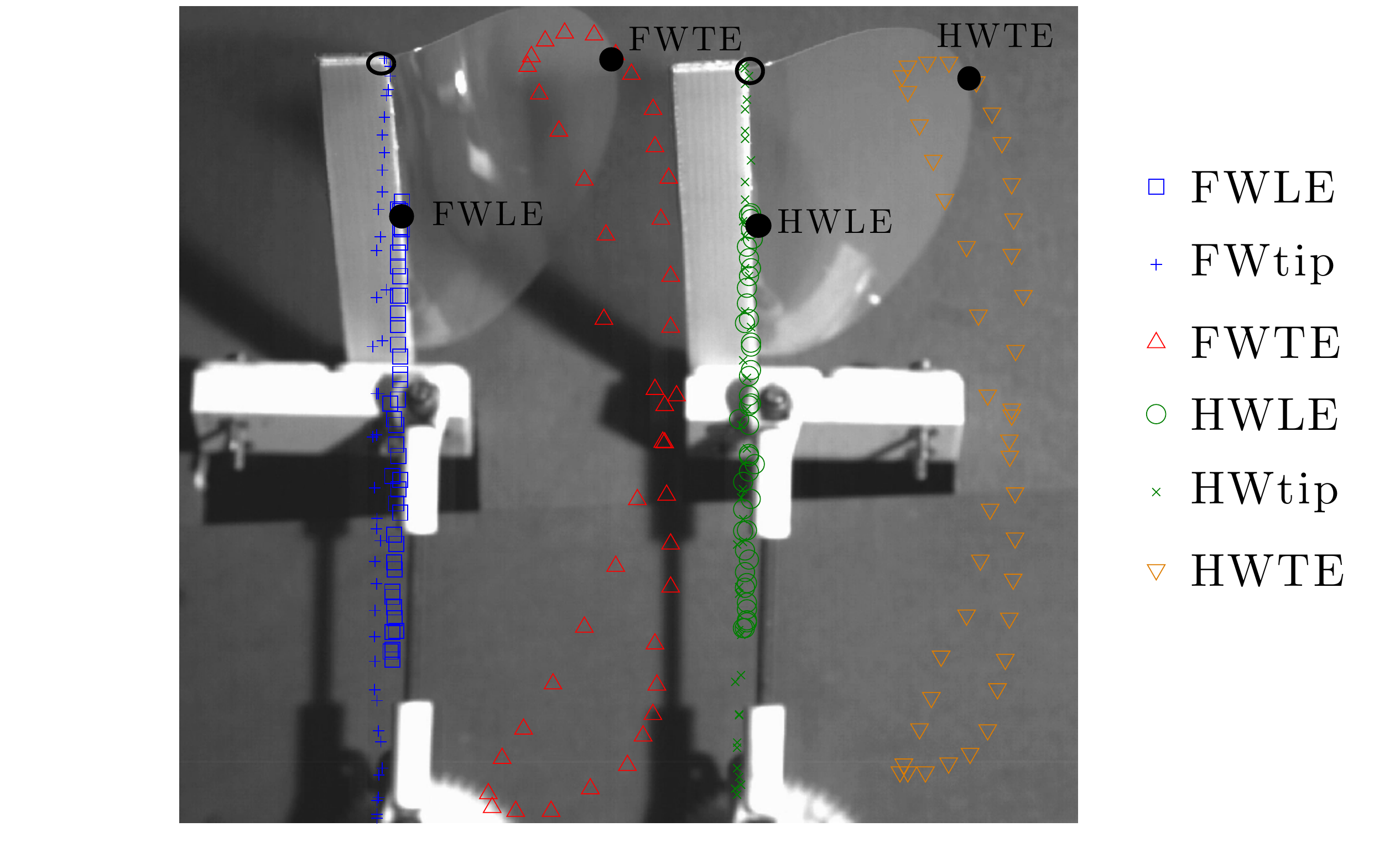}
  \caption{Tracking the wing motion. Example of in-phase flapping at $f=23$Hz. Black circles are the points tracked on each wing for the leading and trailing edges (FWLE and FWTE, correspond to the forewing, while HWLE and HWTE correspond to the hindwing). The open circles are the positions of the win tips tracked.}
\label{fig_tracking}
\end{figure}

\begin{figure}[t]
\centering
\includegraphics[width=0.48\linewidth,trim=15 0 25 0,clip]{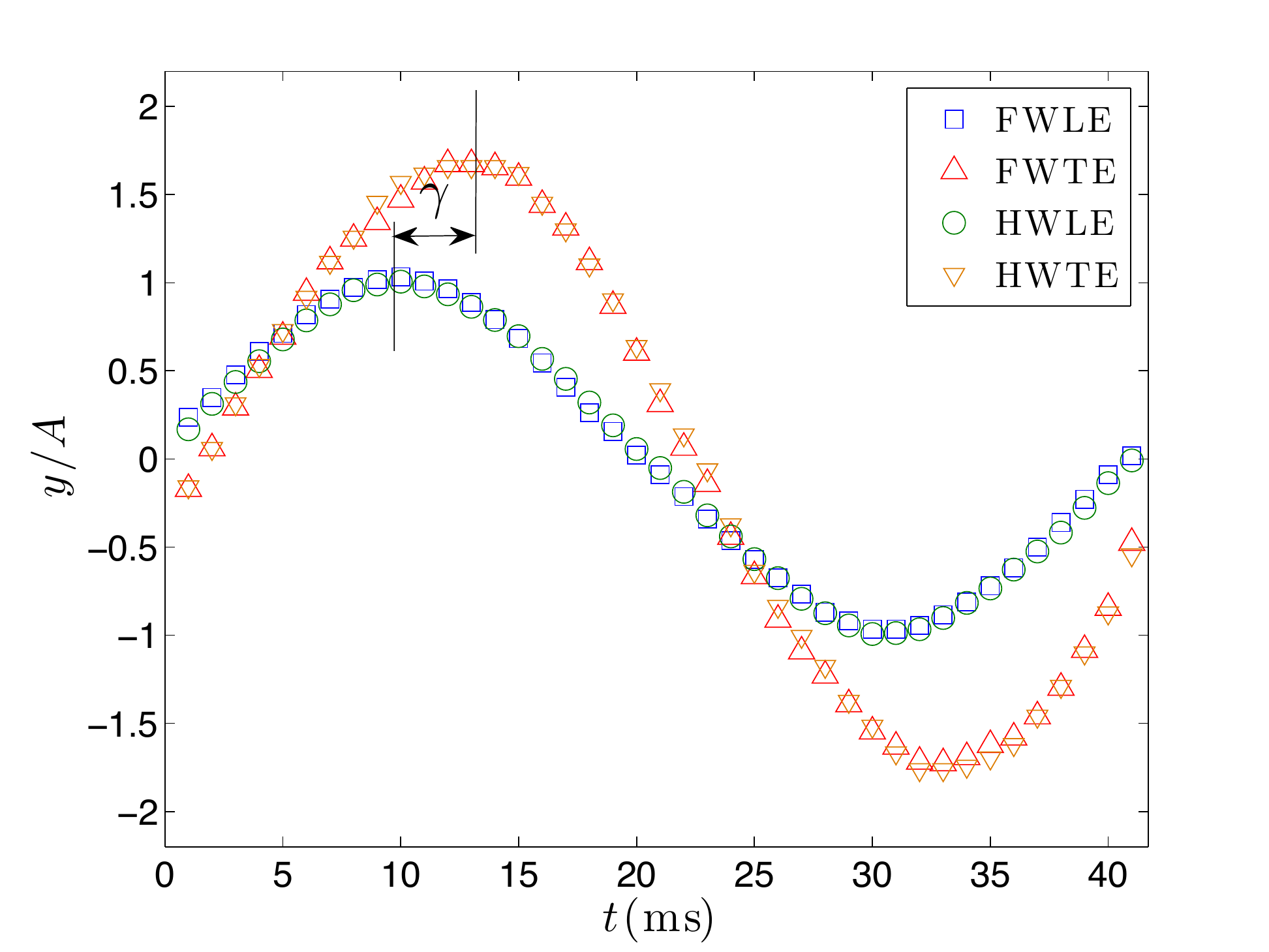}
\includegraphics[width=0.48\linewidth,trim=15 0 25 0,clip]{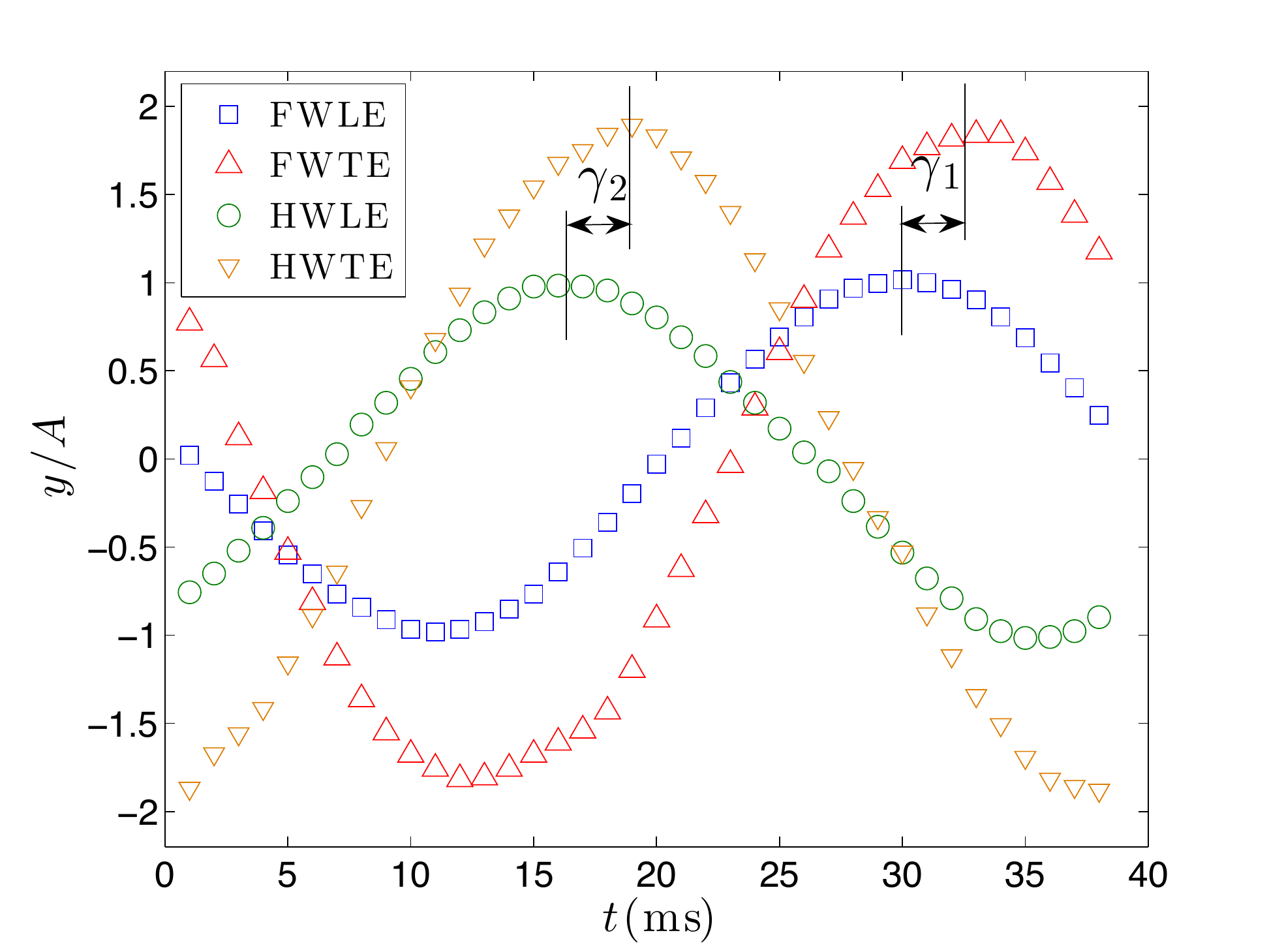}
\includegraphics[width=\linewidth]{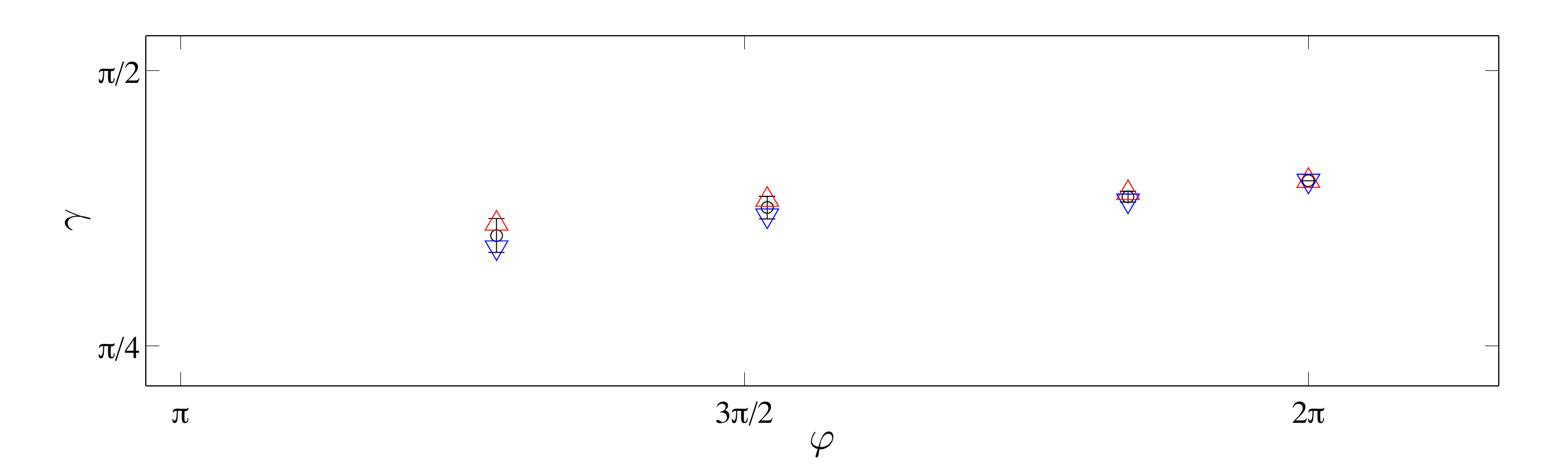}
  \caption{Time series of the $y$-position of the leading and trailing edges of both wings for $f=23$Hz and (a) $\varphi=0$ and (b) $1.3\pi$. The leading-edge-trailing-edge phase lag $\gamma$ is shown schematically. In the $\varphi=0$ case, $\gamma_1=\gamma_2\equiv\gamma$. The $y$-axis is rendered dimensionless using $A$, the peak-to-peak amplitude swept by the leading edge at mid-chord. (c) Trailing-edge-leading-edge phase lag $\gamma$ for different forewing-hindwing phase lags. The two different markers correspond to the forewing $\gamma_1$ (\textcolor{red}{$\bigtriangleup$}) and hindwing $\gamma_2$ (\textcolor{blue}{$\bigtriangledown$}).}
\label{fig_timeseries}
\end{figure}

Figure \ref{fig_tracking} presents the tracking in time of the $(x,y)$ positions of the leading and trailing edges of each wing at the point of maximum chord length (black circles, FWLE and FWTE denoting forewing leading and trailing edge, respectively, while HWLE and HWTE correspond to the equivalent points on the hindwing). Additionally, the tip of each wing is also tracked (empty circles). The latter is useful to minimize the measurement error in the leading edge amplitude of motion since it has a larger swept amplitude and the relative error is thus smaller. The time-series of the positions of these points are shown in figure \ref{fig_timeseries} (a) and (b) for two different values of $\varphi$. The main observation here is the trailing-edge-leading-edge phase lag $\gamma$ for each wing, which has been reported in \cite{ramananarivo2011} to be a crucial element of the propulsive performance of flexible flapping wings. The measured value of $\gamma$ for several forewing-hindwing phase lags is shown in figure \ref{fig_timeseries} (c). A slight decrease is observed for $\gamma$ when $\varphi$ decreases from $2\pi$ to $1.3\pi$, i.e. when the systems goes from in-phase flapping to a configuration with the hindwings leading by approximately a third of a period, but the effect is very weak.

\section{Discussion}
Before discussing the effect of the forewing-hindwing phase lag we start with a comment on the role of the flapping frequency. The existence of an optimal flapping frequency for the present setup (as is evident in Fig. \ref{fig_contours}) is related to the effect of wing compliance on the propulsive performance of the flapping wings, a question that has been widely discussed in the literature \cite[][]{shyy2010,spagnolie2010,masoud2010,thiria2010,ramananarivo2011,kang2011}. For flexible wings flapping in air, where the mass ratio of the wing with respect to the surrounding fluid is high, the main bending motor is the wing inertia \cite[][]{daniel2002,thiria2010}. Increasing the flapping frequency leads to an increased deformation of the wing, which is useful in terms of propulsive performance up to a certain point where the effective lifting surface is diminished and thrust production drops \cite[][]{ramananarivo2011}. An interesting point is that the optimal frequency $f_{opt}$ that can be estimated by inspection of Fig. \ref{fig_contours} is different depending on whether one considers maximum cruising speed ($f_{opt}\sim 24$ Hz) or maximum efficiency ($f_{opt}\sim 21$ Hz). This can be explained by the fact that the consumed power, which enters the quotient defining efficiency (Eq. \ref{eff}) in the denominator, increases monotonically with an increasing frequency. This will thus shift the location of the peak in the efficiency that corresponds to the aerodynamic power peak.

In order to focus on the effect of the forewing-hindwing phase lag, a fixed frequency can be chosen. In Fig.~\ref{fig_results},  for each of the two frequencies represented, the same trends are observed in the cruising speed and the consumed power and, as shown in the insets, the frequency dependence is well explained using the ``elasto-inertial'' dimensionless variables $u$ and $p$ defined above. Now, with respect to the phase lag $\varphi$, two main points appear clearly: on the one hand, the fastest flying performance is found around in-phase flapping (i.e $\varphi=0$), while around anti-phase flapping ($\varphi=\pi$) this cruising speed is the lowest. We may note that the curve is not symmetric with respect to $\varphi=0$, since it diminishes only slightly (one could consider a plateau) between $\varphi=0$ and $\pi/2$, whereas it drops more rapidly when the hindwing leads (i.e. when going from $\varphi=0=2\pi$ towards $\varphi=3\pi/2$). The consumed power curve on the contrary has two maxima (around $\varphi=0$ and $\pi$) and two minima around ($\varphi=\pi/2$ and $3\pi/2$). In-phase and anti-phase flapping being the most expensive can be explained by a simple inertial argument since during these configurations the motor has to accommodate the acceleration/deceleration of both pairs of wings at the same time, contrary to the intermediate phase lags $\varphi=\pi/2$ and $3\pi/2$ where the deceleration of one pair of wings occurs while the other pair is accelerating, hence redistributing the load on the motor.

The shape of the cruising speed and consumed power curves determines the location of the maximum observed in the efficiency contours in Fig. \ref{fig_contours} being around $\varphi=\pi/2$. A secondary maximum can be expected around $\varphi=3\pi/2$ (a zone of the parameter space that was not fully explored in the present experiments). We can now compare the optimum phase-lags that lead to peaks of cruising speed and of efficiency and comment on their physical origin: while the maximum cruising speed is observed for phase-lags between zero and $\pi/4$, the optimum phase-lags in terms of efficiency are around $\pi/2$. The former are ruled solely by the aerodynamics, where the performance of different kinematic patterns will have to be analysed considering for instance the interaction between hindwing and the vortex structures shed by the forewing. A wake capture process of this sort has been proposed by \cite{kolomenskiy2013} as a possible explanation for the large propulsive force found at $0.75\pi$ in their 2D numerical simulations.\footnote{Note that in \cite{kolomenskiy2013} the phase lag is defined with a negative sign with respect to $\varphi$ used here so that $0.75\pi$ here corresponds to their $1.25\pi$.} Concerning the optimum phase lag in terms of efficiency requires on the other hand considering the power consumption, which is not only correlated to aerodynamics, but has a large contribution determined by solid inertia, as we have mentioned in reference to the power plot in Fig. \ref{fig_results}b.

The effect of the modulation of $\varphi$ in terms of aerodynamics, as mentioned previously, will be intrinsically related to the roles of the distance $d$ between the two wing pairs and of the deformation kinematics. In this paper we have fixed $d$ and considered a single flexibility in order to explore the possible roles of the wing deformation in the propulsive performance. Considering our previous studies on the effect of flexibility with a two wing flyer \cite[][]{thiria2010,ramananarivo2011}, where the trailing-edge-leading-edge phase lag $\gamma$ was shown to be a crucial parameter to determine performance, we attempted here to monitor modifications in $\gamma$ as a function of $\varphi$ (see the analysis of wing kinematics in Figs. \ref{fig_tracking} and \ref{fig_timeseries}). We picked a range of $\varphi$ (decreasing from $2\pi$ to $1.3\pi$) where a clear deterioration of the flying performance is observed in figure \ref{fig_results} as the hindwing starts leading the forewing. The observations on the trailing-edge-leading-edge phase lag $\gamma$ are however not conclusive. A slight decrease of $\gamma$ is indeed observed while $\varphi$ goes from $2\pi$ to $1.3\pi$, which at these frequencies could explain a diminishing thrust performance, but the effect is too weak and the present results do not permit to give a thorough and quantitative confirmation of the effect if it exists. 

\section{Conclusions}

We have shown that a four-winged flapping flyer in a cruising regime does present different optimal forewing-hindwing phase lags depending on whether one would want to tune the system for maximum cruising speed or minimum energy expenditure. These results are in accordance with previous studies in hovering configurations by \cite{wang2007,usherwood2008}, hinting that the mechanisms described here should be robust elements to consider in any aerodynamic model: one the one hand wing inertia is a major player in the power expenditure, while on the other hand the thrust production is ruled by the aerodynamics around the flexible wings. 

A full flow field reconstruction around the wings is certainly desirable to compare different points in the parameter space with different performances and clearly identify the aerodynamic mechanisms at play. Experimentally this can be challenging, and numerical simulations such as a 3D version of the fluid-structure simulations of \cite{kolomenskiy2013} can be an interesting option to define robust aerodynamic models.

Other issue that remains not fully understood is the effect of the forewing-hindwing separation $d$ as a parameter independent of the forewing-hindwing phase lag which will have a different role depending on whether the system is hovering or cruising. This point will be of particular importance when going beyond the ``steady'' regimes of hovering or cruising and into the study of transient regimes. These bring indeed a vast set of open questions related to the unsteadiness ---like for instance the multi-body dynamics of manoeuvring--- and where the wings are to be analysed as part of a full system accelerating, performing sharp turns \cite[e.g.][]{bergou2010} or taking off \cite[e.g.][]{bimbard2013}.

\bigskip
{\noindent\textbf{Acknowledgements} We thank all the people at the PMMH workshop for their help with the experimental setup, in particular D. Pradal who conceived and built most of its parts. We thank also X. Benoit-Gonin, M. Vilmay and G. Lemoult for the LabView interfacing,  and S. Ramananarivo, V. Raspa, D. Kolomenskiy and G. Spedding for useful discussions. This work was supported by the French National Research Agency through project No. ANR-08-BLAN-0099 and by EADS Foundation through project "Fluids and elasticity in biomimetic propulsion". M. C. acknowledges financial support from CONACyT and UNAM, M\'exico. P. J. acknowledges support from the Charpak Internship Program.}

\bibliographystyle{springer}
%\bibliography{mite4ailes}

\end{document}